\documentclass{article}
\usepackage{spconf,amsmath,graphicx}
\usepackage{subfigure}
\usepackage{tikz}
\usetikzlibrary{arrows,snakes,backgrounds,patterns,matrix,shapes,fit,calc,shadows,plotmarks}
\usepackage{colortbl,booktabs}
\usepackage{multirow}
\usepackage{subfigure}
\usepackage{verbatim}
\usepackage{array}
\usepackage{algorithm}
\usepackage{xcolor}
\usepackage{algorithmic}
\usepackage{bbding}
\usepackage{amssymb}
\usepackage{amsthm}
\newtheorem{theorem}{Theorem}

\newcommand{\PreserveBackslash}[1]{\let\temp=\\#1\let\\=\temp}
\newcolumntype{C}[1]{>{\PreserveBackslash\centering}p{#1}}
\title{Solving the Capsulation Attack against Backdoor-based Deep Neural Network Watermarks by Reversing Triggers}
\name{Fang-Qi Li, Shi-Lin Wang, Yun Zhu}
\address{\texttt{ \{solour\_lfq,wsl\}@sjtu.edu.cn}\\
School of Electronic Information and Electrical Engineering, Shanghai Jiao Tong University.}
\begin{document}
%\ninept
%
\maketitle
\begin{abstract}
 Backdoor-based watermarking schemes were proposed to protect the intellectual property of artificial intelligence models, especially deep neural networks, under the black-box setting.
Compared with ordinary backdoors, backdoor-based watermarks need to digitally incorporate the owner's identity, which fact adds extra requirements to the trigger generation and verification programs. 
Moreover, these concerns produce additional security risks after the watermarking scheme has been published for as a forensics tool or the owner's evidence has been eavesdropped on. 
This paper proposes the capsulation attack, an efficient method that can invalidate most established backdoor-based watermarking schemes without sacrificing the pirated model's functionality.
By encapsulating the deep neural network with a rule-based or Bayes filter, an adversary can block ownership probing and reject the ownership verification.
We propose a metric, CAScore, to measure a backdoor-based watermarking scheme's security against the capsulation attack.
This paper also proposes a new backdoor-based deep neural network watermarking scheme that is secure against the capsulation attack by reversing the encoding process and randomizing the exposure of triggers.

\end{abstract}
\section{Introduction}
Watermark has been considered as a promising technique in protecting the copyright of artificial intelligence products, especially deep neural networks (DNN).
Based on the type of access to the suspicious DNN, watermarking schemes are classified into white-box DNN schemes and black-box DNN ones~\cite{my1}.
White-box DNN watermark schemes encode the owner's identity information into the network's weights and parameters, whose revealing is possible only after the pirated DNN can be accessed as a white-box.
There have been various studies concerning the location of watermarking, the encoding and decoding formulation~\cite{my2}, the neuron permutation attack and the defense method~\cite{my3}, etc.
Meanwhile, the black-box DNN watermarks that are capable of verifying the ownership even if the suspicious DNN can be accessed as no more than a black-box are considered as the more practical choice.

Since the pioneering works, black-box DNN watermarks have been implemented through DNN backdoors~\cite{my4}.
This is because, in the black-box setting, the input-output relationship is the only source of information, whereas the backdoor is essentially defined as specialized input-output pairs.
By encoding identity information into specific backdoors, some schemes established ownership protection for the black-box setting.
Recent efforts have been devoted to defending the backdoor-based watermark from attacks including blind tuning~\cite{my5}, anomaly detection~\cite{my6}, etc.

It is remarkable that despite these results, there are several fundamental differences between the ordinary DNN backdoor and the black-box DNN watermarks.
For example, in the ownership verification scenario, it is necessary that the response reflects the owner's identity in an unambiguous, unforgeable, and provable manner. 
Therefore, it is necessary that generating triggers is expensive for adversaries. but the difficulty in generating a backdoor trigger is never an aspect of interest for backdoor algorithms.
As the cost of generating backdoor triggers is hard to measure (not to mention proving the difficulty of generating them), researchers have proposed to use a collection of multiple triggers and encode the ownership information into their correlations, in which case the level of security grows with the number of triggers~\cite{my7}.
Unfortunately, these schemes ignore the fact that an adversary knowing the encoding scheme or being given multiple triggers has the ability to distinguish triggers from normal queries and devastate the ownership proof. 
Notice that, unlike a backdoor attack, this knowledge should be made available to every party including the adversary (the watermarking scheme is inapplicable if only few parties can run the scheme), and blocking triggers need not to tune the DNN in the black-box setting.
These observations imply more requirements for triggers and their encoding methods for backdoor-based DNN watermarking schemes, which go beyond the scope of traditional studies on DNN backdoors.

To confront the deficiencies of current black-box DNN watermarking schemes, we incorporate the previous observations and formulate a new threat, the \textbf{Capsulation Attack}, by encapsulating the stolen DNN with an extra filter to invalidate triggers and thence ownership queries.
Then we derive the necessary conditions for a secure black-box DNN watermarking scheme w.r.t. this attack.
We also provide an ownership verification protocol that minimizes the security risk under the black-box setting.
The contributions of this paper are:
\begin{itemize}
\item We propose the capsulation attack against black-box DNN watermarks.
It can easily devastate current schemes with little cost and turns out to be a practical threat to copyright regulation.
A new security metric CAScore is defined according to this attack.
\item A new black-box DNN watermarking scheme is proposed to establish the ownership protection under the capsulation attack by reversing the trigger encoding and ownership verification process.
We show that the ownership integrity can be sufficiently preserved by a delicate encoding scheme without resorting to any advanced cryptological primitives even if the adversary is very knowledgable.
\end{itemize}

\section{Related Works and Preliminaries}
A backdoor of a DNN is a pair of input (known as the trigger) and output with a mapping relationship deviates from the network's normal functionality~\cite{my8}.
For example, a DNN with a backdoor might recognize dogs as cats or miscalculate the statistics of inputs with a specific noise.
As a result, researchers have been using the backdoor as evidence for ownership proof.
By incorporating the identity information into the backdoor and evoking it from the black-box DNN, the ownership can be verified~\cite{my4}.

\begin{table*}[!t]
\centering
\caption{Comparisons between the backdoor attack and the backdoor-based DNN watermarking scheme.}
\begin{center}
\begin{tabular}{c|c|c|c}
\toprule
\textbf{Scheme} & \textbf{Trigger pattern} & \textbf{No. of triggers} & \textbf{Threat Model} \\
\hline
Backdoor attack & Arbitrary/Zero-bit & 1 & White-box tuning\\
%\hline
Backdoor-based watermark & Requires encoding/Multi-bit & $\geq 1$ & Black-box filtering \\
\bottomrule
\end{tabular}
\label{table:1}
\end{center}
\end{table*}

Early backdoor-based watermarks are zero-bit, i.e., no digital information is encoded into the backdoor, so they cannot be presented as legal evidence.
These schemes include stamped triggers and out-of-dataset triggers~\cite{my9}.
Some works focused on persistency and robustness rather than unambiguity and security also use zero-bit triggers~\cite{my10}.
Recent works usually resort to images with outranged pixels~\cite{my11}, images with encoded stamps or invisible perturbations~\cite{my12}, etc.
%To establish unforgeable ownership proof, chained trigger was designed so the probability of forging the ownership evidence declines with the number of triggers.

The differences between the backdoor attack and the backdoor-based DNN watermarking scheme is significant, we list some of them in Table~\ref{table:1}.
The trigger pattern and number of triggers differ due to the difference in purposes between two schemes.
Regarding the threat model, we emphasize that backdoor attacks cannot modify the architecture of the DNN, and the detection and erasion of backdoors are always assumed to take place under the white-box setting~\cite{my5}.
Yet such tunings uniformly lack theoretical basic and remain empirical.
In fact, a party with no knowledge of the training dataset cannot discriminate backdoors from the DNN's normal performance, while an adversary holding the comprehensive knowledge is not motivated to perform backdoor clearance.
The case becomes more complex for backdoor-based watermarks.
If some party claims to have incorporated a backdoor into service of another party with only black-box access, then such a claim is plausible only if these two parties prove beforehand that they do not conspire.  This is because inserting/canceling certain backdoors for a black-box service can be trivially implemented through modules independent of the backend DNN.
%This is a latent dilemma in using backdoors for black-box ownership verification.
Even if the party who issues an ownership claim and the party who provides the suspicious service share no interest, ordinary zero-bit triggers can never be the legal evidence.
The reason behind is that the ownership proof is no more reliable than the unforgeability of the backdoor, so if the backdoor can be easily generated (as all zero-bit triggers that can be produced simply from black-box adversarial learning) then the ownership can be qucikly forged.
Being confronted by this threat, researchers have been resorting to extra backdoor coding schemes.
The intuition is that although creating one trigger is almost always easy, but the difficulty of creating a chain of triggers with correlated patterns and labels grows with the length of the chain and the probability of piracy becomes negligible when the chain is long enough~\cite{my7}.

In general, an unforgeable backdoor-based DNN watermarking scheme can be formulated as a quintet $\texttt{WM}=\langle \texttt{KeyGen},\texttt{Encode},T(\cdot),L(\cdot),\mathbf{N} \rangle$, where
\begin{itemize}
\item \texttt{KeyGen} is the module for identity generation, it returns an identity key:
$$\texttt{Key}\leftarrow\texttt{KeyGen}(\mathbf{N}),$$
\item \texttt{Encode} is the module for identity encoding, it maps an identity key to a series of $N$ codes, each with length $R$ ($N$ and $R$ are parts of $\mathbf{N}$):
$$\left\{\mathbf{c}_{n} \right\}_{n=1}^{N}\leftarrow\texttt{Encode}(\texttt{Key}).$$
\item $T(\cdot)$ is the trigger generator that maps a code $\mathbf{c}$ into a trigger for the DNN to be protected:
$$T(\mathbf{c}_{n})=\mathbf{t}_{n}.$$
\item $L(\cdot)$ is the label generator that maps a code $\mathbf{c}$ into the prediction for $T(\mathbf{c})$:
$$L(\mathbf{c}_{n})=l_{n}.$$
\item $\mathbf{N}$ is the security parameter tuple that determines the length of the identity key/codes and the number of triggers.
\end{itemize}

\begin{table*}[!t]
\centering
\caption{Settings for the capsulation attack.
\checkmark denotes that the adversary knows the component, $\times$ denotes that the adversary has no access to the component.
The knowledge on \texttt{KeyGen}, \texttt{Encode}, and $L(\cdot)$ is assumed to be always available (they can be replaced by any pseudorandom function with correct input-output spaces without loss of generality).
When both $T(\cdot)$ and \texttt{Key} are available, all triggers can be computed.
When $T(\cdot)$ is unavailable, triggers cannot be produced from \texttt{Key} so the knowledge on it is insignificant.}
\vspace {-2.5mm}
\begin{center}
\begin{tabular}{c|c|c|||c|c|c|c}
\toprule
\texttt{KeyGen} & \texttt{Encode} & $L(\cdot)$ & $T(\cdot)$ & \texttt{Key} & A subset of $\left\{\mathbf{t}_{n} \right\}_{n=1}^{N}$ & \textbf{Type of the filter} \\
\hline
\checkmark & \checkmark & \checkmark & \checkmark & \checkmark & \checkmark/$\times$ & Rule-based filter \\
%\hline
\checkmark & \checkmark & \checkmark & \checkmark & $\times$ & \checkmark & Bayes filter/Rule-based filter \\
%\hline
\checkmark & \checkmark & \checkmark & \checkmark & $\times$ & $\times$ & Bayes filter/Rule-based filter \\
%\hline
\checkmark & \checkmark & \checkmark & $\times$ & \checkmark/$\times$ & \checkmark & Bayes filter \\
%\hline
\checkmark & \checkmark & \checkmark & $\times$ & \checkmark/$\times$ & $\times$ & Inapplicable \\
\bottomrule
\end{tabular}
\label{table:2}
\end{center}
\end{table*}

\begin{figure}[!t]
\centerline{\includegraphics[width=3.15in]{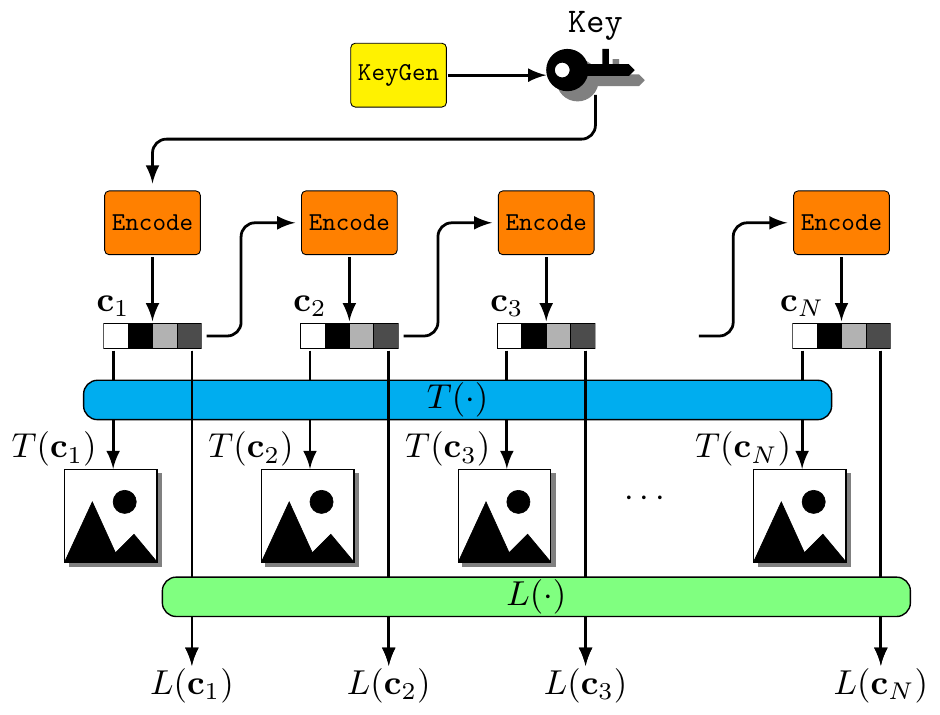}}
\caption{The workflow of a backdoor-based DNN watermarking scheme. The \texttt{Encode} works in a chain so increasing the number of triggers need not structural alternation.}
\label{figure:1}
\end{figure}

To watermark its DNN, the owner runs \texttt{KeyGen}, \texttt{Encode}, generates $N$ backdoors using $T(\cdot)$, $L(\cdot)$ as shown in Fig.\ref{figure:1}, and tunes the DNN with the backdoor dataset $\left\{(T(\mathbf{c}_{n}),L(\mathbf{c}_{n})) \right\}_{n=1}^{N}$.
To prove its ownership to a third party, the owner only needs to submit $\texttt{Key}$, where the rest components are available for any party for authorization.
Then the third party reconstructs backdoor triggers from \texttt{Key} and examines whether the outputs of the suspicious service are in accordance with the predicted labels.
The security of this scheme relies on this observation: the probability that an irrelevant identity key declines exponentially in $N$, so an ambiguity attack can hardly succeed.
It is also remarkable that $L(\cdot)$ need not be task-dependent.
The label encoder only has to introduce a certain equivalence relationship among codes and this relationship can be retrieved even if an adversary shuffles the labels, incorporates or deletes labels, or adds its own downstream modules as in language processing.
We emphasize that the quintet \texttt{WM} should be assumed as available for every party to establish a forensics system since the permission of arbitrary encoders or generators results in a direct breach of the ownership's unforgeability.

Unfortunately, this setting is miserably insufficient for ownership protection in the field.
An adversary with the knowledge of the backdoor triggers (in particular, the knowledge on $T(\cdot)$) can filter triggers to deny the ownership proof.
This filter is different from neural cleanse~\cite{my13} or adversarial tuning where we try to erase a backdoor from a DNN without modifying its architecture.
As long as the DNN is assumed to be hidden as a black-box, the adversary is free to add extra modules rather than tuning the DNN.
Meanwhile, we do not have to consider blind tuning schemes that are designed to cancel backdoors with unknown patterns, since hiding the triggers' distribution results in more threats as zero-bit triggers (an adversary can then claim an arbitrary sample as its ownership proof to boycott legal services) and is unacceptable for ownership protection.
Hitherto, studies on this aspect have been very limited except for a new backdoor embedding method~\cite{myexpo} and penetrative perturbations~\cite{my6}.
Neither an analytic bound nor a theoretical framework has been proposed to formally address this threat.

\section{The Threat Model}
\begin{figure}[!t]
\centerline{\includegraphics[width=3.15in]{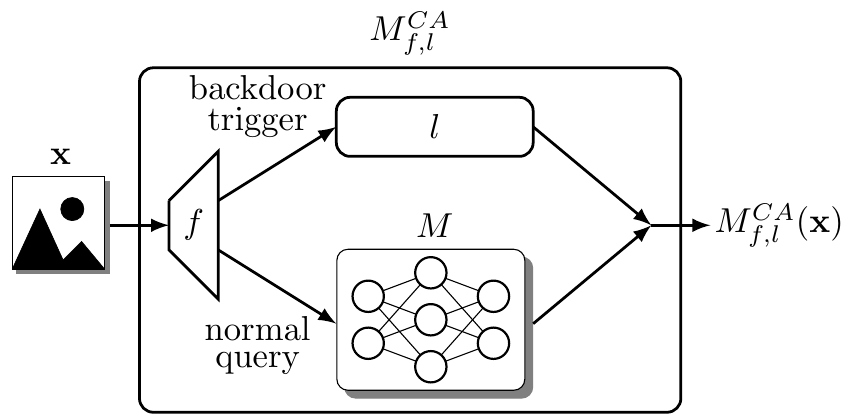}}
\caption{The capsulation attack $M^{\text{CA}}_{f,l}$.}
\label{figure:2}
\end{figure}

We consider the scenario where an adversary has learned the watermarking scheme, pirated the DNN denoted as $M$, and deployed a commercial service upon it as a black-box.
The only difference from the naive black-box setting is that the adversary adds a filter $f$ in front of the DNN to detect whether the input $\mathbf{x}$ is an trigger for ownership query or not.
For an ownership query, the service returns a fake label $l(\mathbf{x})$.
This prototype operates as shown by Algo.1 and is visualized in Fig.\ref{figure:2}.

\begin{algorithm}[htbp]
\caption{$M^{\text{CA}}_{f,l}$: Capsulation attack for model $M$.}
\label{algorithm:1}
\textbf{Input}: The pirated DNN model $M$, input $\mathbf{x}$. \\
\textbf{Parameter}: The filter $f$, the fake label generator $l$.\\
\textbf{Output}: $M^{\text{CA}}_{f,l}(\mathbf{x})$
\begin{algorithmic}[1] %[1] enables line numbers
\IF {$f(\mathbf{x})==1$ ($\mathbf{x}$ is a backdoor trigger)}
\STATE \textbf{return} $l(\mathbf{x})$.
\ELSE
\STATE \textbf{return} $M(\mathbf{x})$.
\ENDIF
\end{algorithmic}
\end{algorithm}

Having analyzed the components of a black-box DNN watermarking scheme in general, we demonstrate the different forms of capsulation attacks, in particular the type of filters, under different combinations of knowledge exposure detailed in Table~\ref{table:2}.

The first case is when both $T(\cdot)$ and \texttt{Key} have been exposed.
This occurs when the piracy takes place after the owner has presented its ownership proof once.
In this case, the backdoor can be trivially blocked by a rule-based filter.
The filter $f$ maintains a list of triggers derived from \texttt{Key} and returns 1 if $\mathbf{x}$ equals one of the triggers.
The fake label generator $l$ returns an arbitrary label different from $M(\mathbf{x})$.
This capsulation is sufficient for stealing a DNN after eavesdropping on the ownership proof.

In cases where \texttt{Key} is unavailable but $T(\cdot)$ is too simple, as is the case for most established schemes, the filter does not need the knowledge on \texttt{Key} or exposed triggers.
For example, triggers with outranged components~\cite{my11} or random noises~\cite{my7} can almost always be unambiguously recognized.
Therefore, these schemes are inherently insecure in the field.

Finally, some delicate trigger generators produce triggers whose features cannot be trivially recognized or the complete $T(\cdot)$ is unavailable but some of the triggers have been exposed, then the adversary is encouraged to configure a Bayes classifier as the filter.
Denote the event that an input $I$ is a trigger by $I\sim T$, formally, a Bayes filter is formulated as
\begin{equation}
\label{equation:bayes}
\begin{small}
\begin{aligned}
&\text{Pr}\left\{I\sim T|I=\mathbf{x} \right\}= \frac{\text{Pr}\left\{I=\mathbf{x},I\sim T \right\}}{\text{Pr}\left\{I=\mathbf{x} \right\}}\\
=&\frac{\text{Pr}\left\{I=\mathbf{x}|I\sim T \right\}p_{T}}{\text{Pr}\left\{I=\mathbf{x}|I\sim T \right\}p_{T}+\text{Pr}\left\{I=\mathbf{x}|\neg (I\sim T \right)\}(1-p_{T})},
\end{aligned}
\end{small}
\end{equation}
where $p_{T}=\text{Pr}\left\{I\sim T \right\}$ is the prior probability that the input is an ownership query.
In Eq.~\eqref{equation:bayes}, the conditional probability distributions $\text{Pr}\left\{I=\mathbf{x}|I\sim T \right\}$ and $\text{Pr}\left\{I=\mathbf{x}|\neg(I\sim T) \right\}$ can be computed by utilizing $T(\cdot)$ and collecting normal queries during the service.
In general, the adversary can identify $Q$ triggers by running $T(\cdot)$/observing exposed triggers and collecting another $Q$ normal queries on the fly.
Then it trains a binary classifier $f$ on this dataset consisting $2Q$ samples and deploys it as the filter.

The security against the capsulation attack of a black-box DNN watermarking scheme relies on how well can the adversary classify normal queries and backdoor triggers.
We propose a new metric (\textbf{C}apsulation \textbf{A}ttack \textbf{Score}, CAScore) to quantify this aspect of security
\begin{equation}
\label{equation:1}
\text{CAScore}^{\texttt{WM}}=2*(1-\max_{f}\left\{\text{AUC}(f,\texttt{WM}.T(\cdot),\mathcal{D}) \right\}),
\end{equation}
in which $\text{AUC}(f,\texttt{WM}.T(\cdot),\mathcal{D})$ is the area under the receiver operating characteristic curve for the binary classification between triggers generated by $T(\cdot)$ in \texttt{WM} and normal samples from the dataset $\mathcal{D}$ implemented by $f$.
The metric $\text{CAScore}^{\texttt{WM}}\in[0,1]$ measures the difficulty in setting a pertinent filter, and hence the robustness of the watermark.
Obviously $\text{CAScore}^{\texttt{WM}}$ can hardly be analytically computed except for some schemes whose triggers' pattern is too evident (so their CAScore is zero).
In general cases, we can only train a finite collection of classifiers to derive an upper bound of this metric as an estimation for comparing the security levels of different schemes.

%\subsection{Challenge: Overwriting Attack}
%The capsulation attack can not only prevent backdoors from being evoked, it can also simulate arbitrary backdoors and hence the ownership proof of any party.
%Adversaries can leverage this property to conduct the overwriting attack and defeat the authentic owner by a forged time-stamp.
%This attack is demonstrated in Fig.~\ref{figure:2}.
%Ownership is indefensible against this overwriting attack in the ideal black-box setting.
%Although the authentic owner might be able to verify its ownership, its time-stamp remains fragile, so it can hardly convince any third party or authorities.

\section{The Proposed Scheme}
\subsection{The motivation}
Given the metric defined in \eqref{equation:1}, we are left with the challenge of designing a trigger generator so that no classifier can unambiguously distinguish triggers from normal queries.
Theoretically, this could only happen when triggers share exactly the same distribution with the normal data so the optimal choice of the trigger set is a subset of $\mathcal{D}$.
This setting opposes the ordinary trigger generation process and calls for a new method to establish digital unforgeability.
Concretely, we adopt a reverse paradigm by encoding triggers into their hash and presenting them in a randomized order.

\begin{figure}[!t]
\centerline{\includegraphics[width=3.15in]{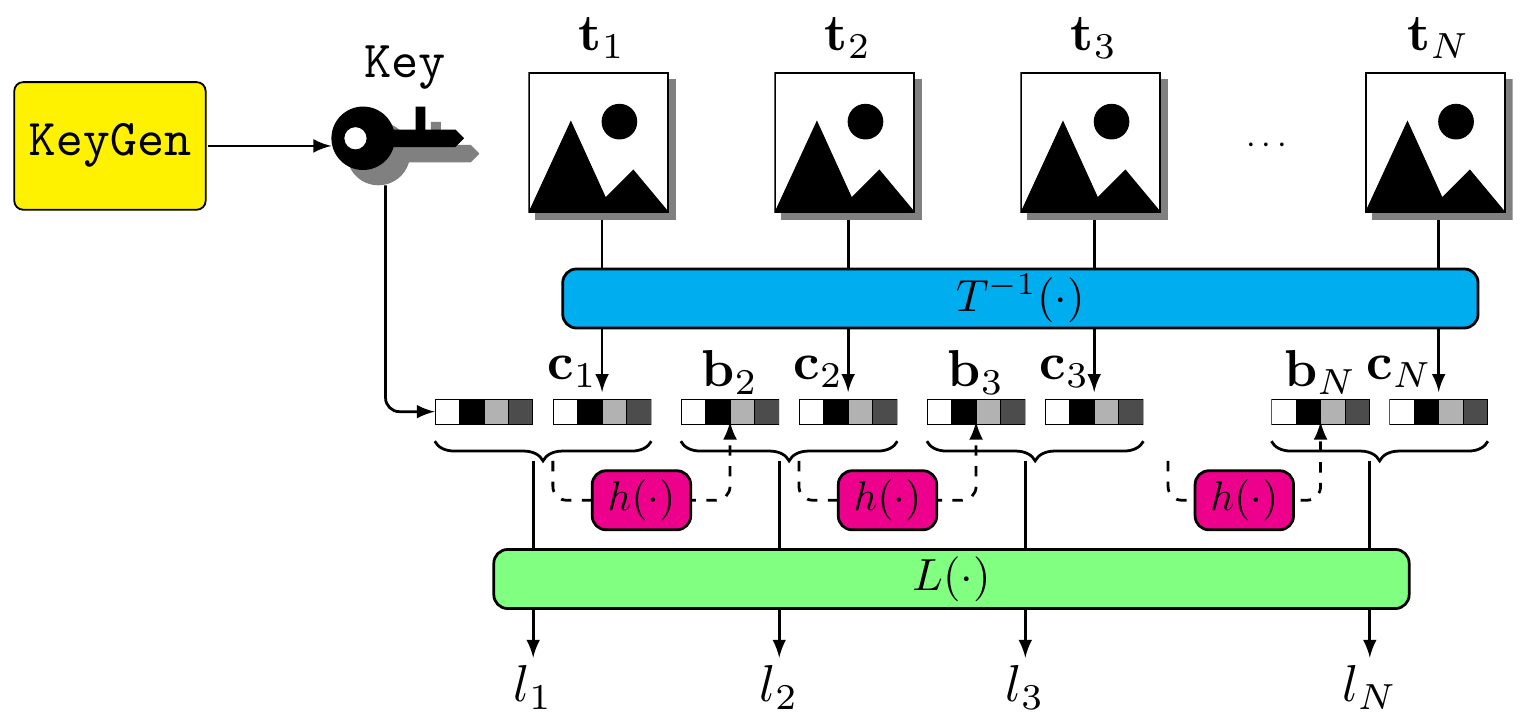}}
\caption{Generating backdoor triggers.}
\label{figure:3}
\end{figure}

\subsection{The reverse-backdoor scheme}
Similar to its predecessors, the reverse-backdoor DNN watermarking scheme, is composed of five elements, $\texttt{WM}=\langle\texttt{KeyGen},T^{-1}(\cdot),h(\cdot),L(\cdot),\mathbf{N} \rangle$, in which $\texttt{KeyGen}$ and $\mathbf{N}$ are identical to the original settings.
\begin{itemize}
\item $T^{-1}(\cdot)$ is a pseudorandom inverse trigger generator (e.g., a hash function) that maps a trigger into a code with length $R$:
$$T^{-1}(\mathbf{t}_{n})=\mathbf{c}_{n}.$$
\item $h(\cdot)$ is a one-way hash function that maps an input with length $2R$ into a code with length $R$.
\item $L(\cdot)$ is the label generator with input length $2R$.
\end{itemize}

To generate backdoors, the owner selects a collection of $N=2^{P}$ samples from the training dataset, denoted as $(\mathbf{t}_{1},\mathbf{t}_{2},\cdots,\mathbf{t}_{N})$.
Then the owner feeds all triggers to $T^{-1}(\cdot)$ and obtains their codes $(\mathbf{c}_{1},\mathbf{c}_{2},\cdots,\mathbf{c}_{N})$.
The label for $\mathbf{t}_{1}$ is assigned as
$$l_{1}=L(\texttt{Key}\|\mathbf{c}_{1}),$$
set $\mathbf{b}_{2}=h(\texttt{Key}\|\mathbf{c}_{1})$.
For $n=2,\cdots,N$, $\mathbf{t}_{n}$'s label is
$$l_{n}=L(\mathbf{b}_{n}\|\mathbf{c}_{n}),$$
with $\mathbf{b}_{n}=h(\mathbf{b}_{n-1}\|\mathbf{c}_{n-1})$ for $n=3,\cdots,N$.
This process is demonstrated in Fig.3.

\begin{figure}[!t]
\centerline{\includegraphics[width=3.15in]{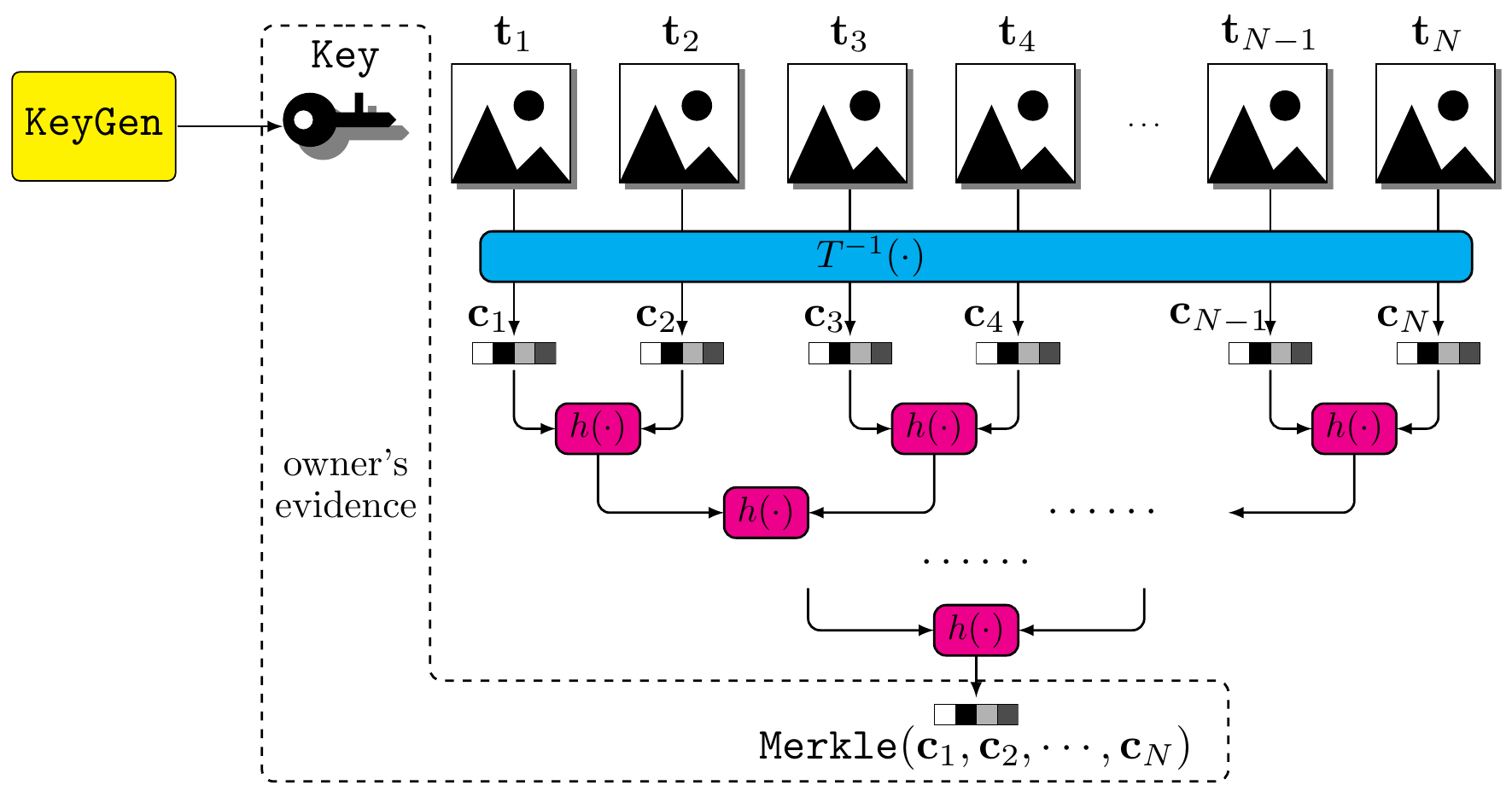}}
\caption{Generating the ownership evidence.}
\label{figure:4}
\end{figure}

The ownership evidence includes the identity $\texttt{Key}$ and the Merkle hash of triggers~\cite{my15} computed by $T^{-1}(\cdot)$ and $h(\cdot)$, formally, $\forall n=1,2,\cdots,N$,
$$\mathbf{c}_{0,n}=\mathbf{c}_{n},$$
then $\forall p=1,2,\cdots,\log_{2}(N)$, $\forall n=1,2,\cdots,\frac{N}{2^{p}}$,
$$\mathbf{c}_{p,n}=h(\mathbf{c}_{p-1,2n-1}||\mathbf{c}_{p-1,2n}),$$
Finally, $\texttt{Merkle}(\mathbf{c}_{1},\mathbf{c}_{2},\cdots,\mathbf{c}_{N})$ is defined as $\mathbf{c}_{P,1}$, this process is shown in Fig.\ref{figure:4}.

To prove its ownership over a suspicious service $M$ to a third party, the owner indicates the previously broadcasted evidence and submits its triggers in a randomized order.
Then the third-party inputs triggers into the suspicious service and sequentially records the service's outputs. 
Finally, the owner gives the order of triggers so the third-party can examine the consistency between the recorded results and the ownership evidence regarding the labels and the Merkle hash.

\subsection{Security analysis}
The unforgeability and unambiguity of the proposed scheme are evident given the one-wayness of $h(\cdot)$, the pseudorandomness of $T^{-1}(\cdot)$, and the chaining construction.
The probability that an adversary succeeds in claiming the ownership over an innocent service from a fixed piece of evidence declines exponentially in $N$, since it is not allowed to modify the triggers afterward.
The proof proceeds as the same line as in~\cite{my7}.
Formally, the probability that all $N$ triggers follow the label defined in \texttt{WM} without modifying the DNN is $C^{-N}$, where $C$ is the size of the domain of $L(\cdot)$.
When the ownership is agreed under an imperfect accuracy then the probability that an ambiguity attack succeeds can be estimated by the following theorem.
\begin{theorem}
If the ownership verification passes with an accuracy threshold of $\tau\in (\frac{1}{C},1)$ then the probability that an ambiguity attack succeeds declines exponentially in $N$.
\end{theorem}
\textit{Proof:} Let $\xi_{n}$ be the random variable whose value is 1 if the adversary's $n$-th trigger's code is consistent with \texttt{WM} and is 0 otherwise.
Let $X=\sum_{n=1}^{N}\xi_{n}$ then
$$\mathbb{E}[X]=\sum_{n=1}^{N}\mathbb{E}[\xi_{n}]=\frac{N}{C},$$
since each trigger is independent.
Now the probability that $X\geq \tau N$ can be bounded by the Chernoff theorem.
\begin{equation}
\nonumber
\begin{aligned}
\text{Pr}\left\{X\geq \tau N \right\}&=\text{Pr}\left\{e^{\lambda X}\geq e^{\lambda \tau N} \right\}\\
&\leq \frac{\mathbb{E}[e^{\lambda X}]}{e^{\lambda \tau N}}\\
&=\left(\frac{\frac{e^{\lambda}}{C}+1-\frac{1}{C}}{e^{\lambda \tau}} \right)^{N},
\end{aligned}
\end{equation}
for any $\lambda>0$.
When $\tau>\frac{1}{C}$, it is always possible to choose $\lambda$ so that $\text{Pr}\left\{X\geq \tau N \right\}$'s upper bound declines exponentially in $N$.
\hfill $\blacksquare$\par
As mentioned in Sec.2, when the backend task is not classification, it is sufficient that $L(\cdot)$ establishes two equivalent classes among all triggers, which case is tantamount to $\hat{C}=2$ and $\hat{N}=N-1$.
These modifications do not change the result of Theorem 1.

The reverse-backdoor scheme enjoys a high CAScore since the triggers follow exact the same distribution as normal queries.
Therefore, an adversary cannot painlessly block ownership queries.
This is an intuitive result since $f$'s overfitting knowledge cannot generalize to further arbitrarily assigned triggers.

Finally, remark that even for an adversary knowing the watermarking scheme, the exposed evidence grants no extra merit in damaging the ownership verification process, since the assigned label of the trigger is computable (for both the adversary and the third party to whom the owner is presenting the ownership proof) only after the order and the previous trigger is given.
By packing the codes of triggers as a Merkle-tree, the adversary cannot infer whether two consecutive inputs are chained triggers or independent queries.
This property is nontrivial and fails to hold if the triggers are submitted in the ordinary order (even in the reverse order) or the label for each trigger is assigned independently. 
In these cases, the adversary can run the watermarking scheme to infer the assigned label of the input and consider the input as an ownership query if the predictions returned by $L(\cdot)$ and $M(\cdot)$ are identical.

\section{Experiments and Discussions}
\subsection{Settings and baselines}
\begin{figure}[htbp]
\centering
\subfigure[Noise.]{
\begin{minipage}[htbp]{0.23\linewidth}
\centering
\includegraphics[width=1.5cm]{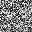}
\end{minipage}
}\subfigure[Wonder Filter.]{
\begin{minipage}[htbp]{0.23\linewidth}
\centering
\includegraphics[width=1.5cm]{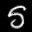}
\end{minipage}
}\subfigure[Stamp.]{
\begin{minipage}[htbp]{0.23\linewidth}
\centering
\includegraphics[width=1.5cm]{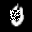}
\end{minipage}
}\subfigure[Steganograhpy.]{
\begin{minipage}[htbp]{0.25\linewidth}
\centering
\includegraphics[width=1.5cm]{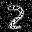}
\end{minipage}
}
\caption{Examples of triggers generated in four baseline watermarking schemes.
}
\label{figure:5}
\end{figure}

To empirically evaluate the proposed method, we conducted experiments on MNIST~\cite{my16} and CIFAR-10~\cite{my17} with the residual network~\cite{my18} as the backbone DNN .
Four backdoor-based watermarking schemes were incorporated as baselines to be compared: \textbf{Noise}, \textbf{Wonder Filter}, \textbf{Stamp}, and \textbf{Steganography}. 
Examples of triggers generated by four schemes are given in Fig.\ref{figure:5}.
\begin{itemize}
\item \textbf{Noise} uses random Gaussian noise as the trigger generator~\cite{my7,my9}.
\item \textbf{Wonder Filter} uses images with outranged pixels as its triggers~\cite{my11}.
\item \textbf{Stamp} adds a stamp onto images as its triggers~\cite{my9}.
\item \textbf{Steganography} exerts an slight perturbation onto images as its triggers~\cite{my6}.
\end{itemize}

\subsection{The CAScore}
\begin{figure*}[htbp]
\centering
\subfigure[$k$ nearest neighbours.]{
\begin{minipage}[htbp]{0.24\linewidth}
\centering
\includegraphics[width=4cm]{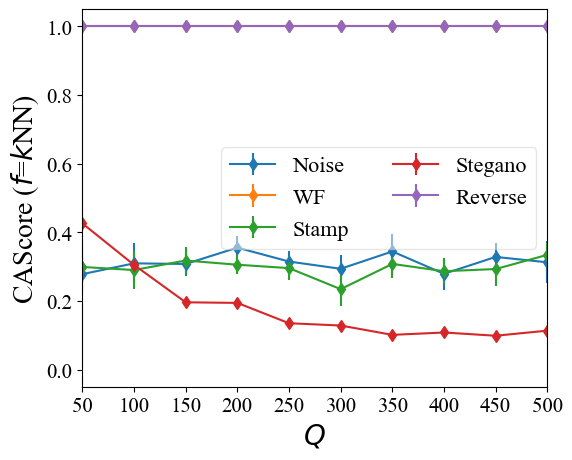}
\end{minipage}
}\subfigure[Naive Bayes.]{
\begin{minipage}[htbp]{0.24\linewidth}
\centering
\includegraphics[width=4cm]{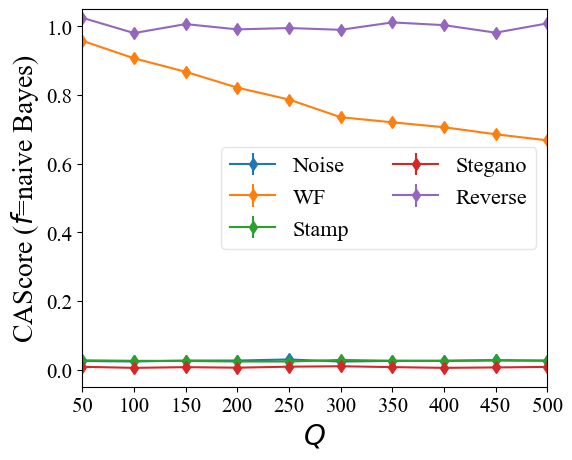}
\end{minipage}
}\subfigure[Logistic regression.]{
\begin{minipage}[htbp]{0.24\linewidth}
\centering
\includegraphics[width=4cm]{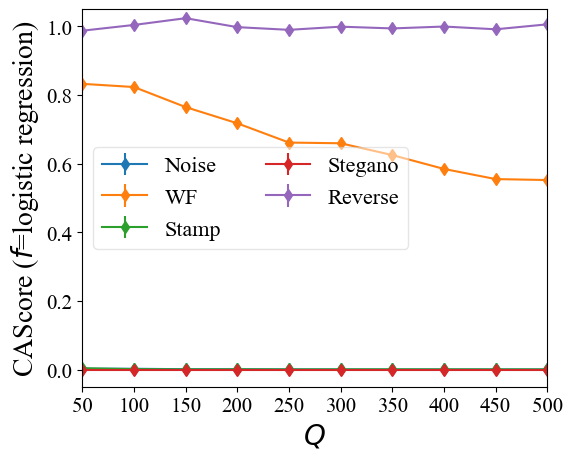}
\end{minipage}
}\subfigure[Neural network.]{
\begin{minipage}[htbp]{0.24\linewidth}
\centering
\includegraphics[width=4cm]{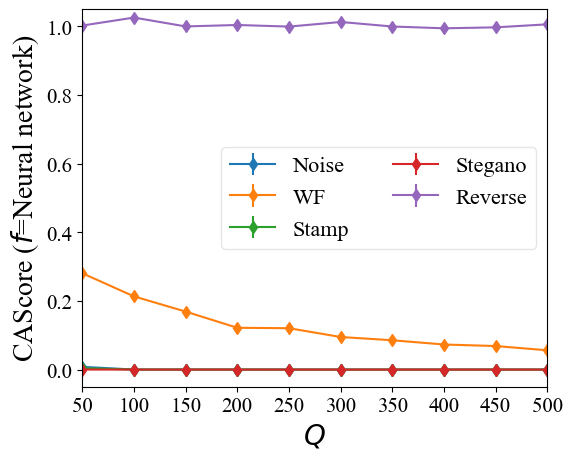}
\end{minipage}
}
\caption{CAScore bounds under different settings.}
\label{figure:6}
\end{figure*}

We first evaluated the security of the studied schemes w.r.t. Eq.\eqref{equation:1}.
As has been discussed, the CAScore can only be upper bounded by exhausting a finite collection of candidate classifiers.
Four basic classifiers were adopted: $k$ nearest neighbors, naive Bayes, logistic regression, and a two-layer neural network.
Under each setting, the adversary was assumed to have obtained $Q$ normal samples during service and $Q$ triggers from the trigger generator.
The CAScores for all combinations among the choices of the watermarking scheme, the classifier, and $Q$ in the range $(50,500)$ were given in Fig.\ref{figure:6}.

\begin{figure}[!t]
\centering
\subfigure[Normal queries, MNIST.]{
\begin{minipage}[htbp]{0.48\linewidth}
\centering
\includegraphics[width=4cm]{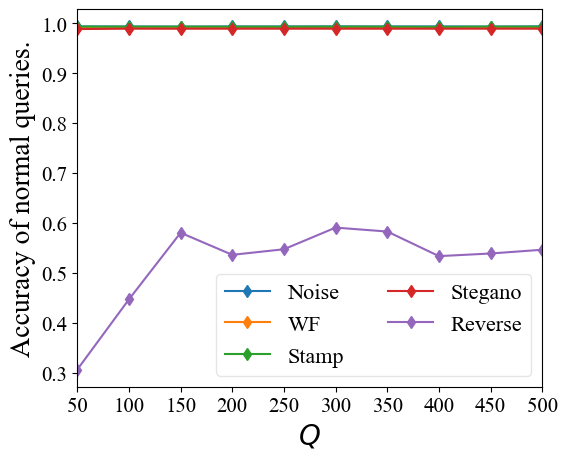}
\end{minipage}
}\subfigure[Ownership queries, MNIST.]{
\begin{minipage}[htbp]{0.48\linewidth}
\centering
\includegraphics[width=4cm]{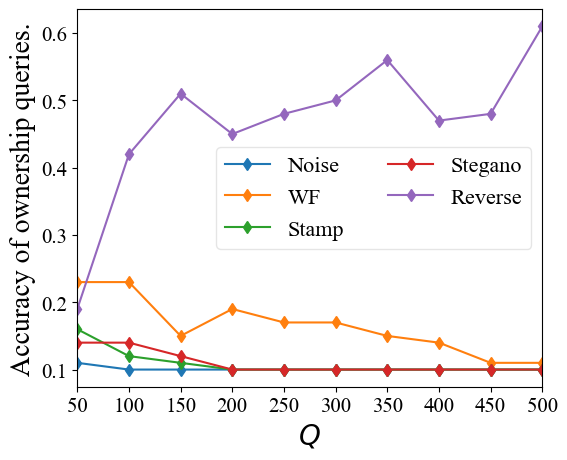}
\end{minipage}
}
\subfigure[Normal queries, CIFAR-10.]{
\begin{minipage}[htbp]{0.48\linewidth}
\centering
\includegraphics[width=4cm]{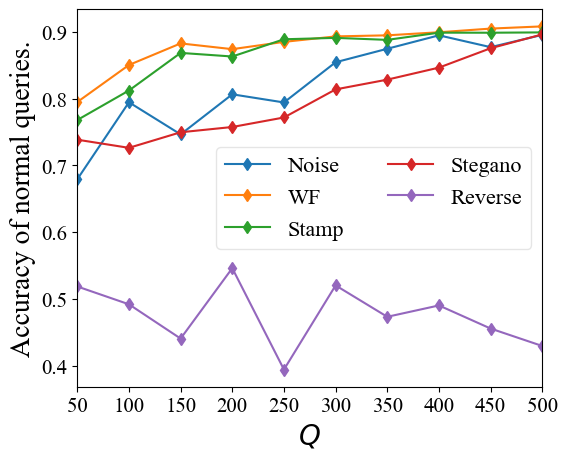}
\end{minipage}
}\subfigure[Ownership queries, CIFAR-10.]{
\begin{minipage}[htbp]{0.48\linewidth}
\centering
\includegraphics[width=4cm]{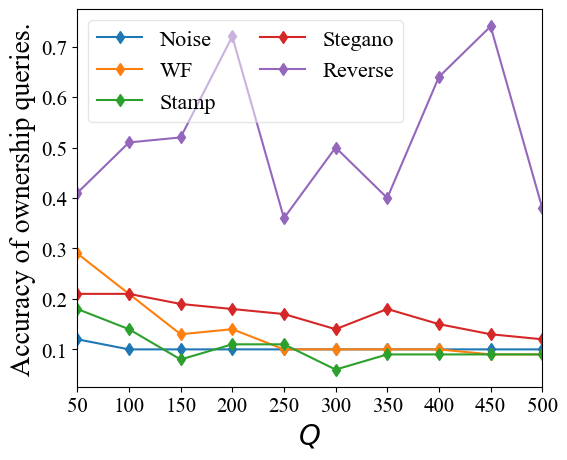}
\end{minipage}
}
\caption{Classification accuracy of different types of triggers under the capsulation attack.
}
\label{figure:7}
\end{figure}

We observed that
1) Complex classifiers had better performance and reduced the upper bound of CAScore.
2) The CAScore bound declined with $Q$, since more information assisted the classifier to better discriminate triggers from normal queries.
3) As summarized in Table~\ref{table:3}, our scheme \textbf{Reverse} enjoyed the optimal CAScore since no classifiers can distinguish triggers from normal queries, even from a theoretical perspective.

Compared with fine-tuning attack families, the capsulation attack is perceptibly capable of exerting more damage and is much easier to commit. 

\subsection{The efficacy of the capsulation attack}
Having equipped with the filter, we considered the case where the adversary applies the capsulation attack proposed in Sec.3.
The adversary adopted the neural network filter with different $Q$s to block ownership queries, and we recorded the classification accuracy of normal inputs and that of backdoor triggers for the capsulated services (where for each scheme, $N=100$ triggers had been incorporated into the DNN), results are shown in Fig.\ref{figure:7}.

As what has been analyzed, schemes with a low CAScore can be invalidated with little expense under the capsulation attack.
This attack had little influence on the service's performance on normal queries when $Q$ was sufficiently large, while that on backdoor triggers was damaged so the ownership was no longer tractable.
Meanwhile, triggers in our scheme survived the filter and the capsulation attack.
If the adversary attempted to distinguish triggers from normal queries and invalidate the former, then it risked the service's accuracy as indicated by Fig.\ref{figure:7}(a)(c).

\begin{table}[!t]
\centering
\caption{The upper bound of CAScore, $Q=500$.}
\begin{center}
\scalebox{0.9}{
\begin{tabular}{c|c}
\toprule
\textbf{Watermarking scheme} & \textbf{CAScore bound} \\
\hline
Noise & 0.00 \\
Wonder Filter & 0.08 \\
Stamp & 0.00 \\
Steganography & 0.06 \\
Reverse & \textbf{0.99} \\
\bottomrule
\end{tabular}
}
\label{table:3}
\end{center}
\end{table}

\subsection{The functionality-preservation evaluation}
Another concern on the reverse-backdoor scheme is that assigning abnormal labels to normal inputs might harm the DNN's performance.
However, we observed that for modern deep models with sufficient redundancy, such harm is negligible.
The DNN's performance on the test set under different configurations is collected in Table~\ref{table:3}, from which we concluded that the performance decline introduced by applying reverse-backdoor was no larger than other candidates.

\begin{table*}[!t]
\centering
\caption{The classification accuracy on MNIST(left) and CIFAR-10(right) under different watermarking configurations (\%).}
\vspace {0.5mm}
\begin{minipage}[c]{0.5\textwidth}
\begin{center}
\scalebox{0.75}{
\begin{tabular}{c|c|c|c|c}
\toprule
\textbf{Scheme} & $N=25$ & $N=50$ & $N=75$ & $N=100$\\
\hline
Noise & $99.31\pm0.08$ & $99.32\pm0.02$ & $99.28\pm0.06$ & $99.23\pm0.10$ \\
%\hline
Wonder Filter & $99.36\pm0.06$ & $99.35\pm0.08$ & $99.28\pm0.02$ & $99.24\pm0.06$ \\
%\hline
Stamp & $99.40\pm0.05$ & $99.34\pm0.04$ & $99.30\pm0.04$ & $99.21\pm0.06$ \\
%\hline
Steganography & $99.35\pm0.03$ & $99.32\pm0.04$ & $99.30\pm0.06$ & $99.25\pm0.08$ \\
%\hline
Reverse & $99.33\pm0.09$ & $99.30\pm0.05$ & $99.26\pm0.02$ & $99.26\pm0.04$ \\
\bottomrule
\end{tabular}
}
\label{table:4}
\end{center}
\end{minipage}%
\begin{minipage}[c]{0.5\textwidth}
\begin{center}
\scalebox{0.75}{
\begin{tabular}{c|c|c|c|c}
\toprule
\textbf{Scheme} & $N=25$ & $N=50$ & $N=75$ & $N=100$\\
\hline
Noise & $92.67\pm0.10$ & $92.66\pm0.08$ & $92.40\pm0.09$ & $92.40\pm0.04$ \\
%\hline
Wonder Filter & $92.63\pm0.06$ & $92.59\pm0.05$ & $92.50\pm0.04$ & $92.47\pm0.04$ \\
%\hline
Stamp & $92.61\pm0.09$ & $92.52\pm0.10$ & $92.41\pm0.08$ & $92.40\pm0.09$ \\
%\hline
Steganography & $92.6\pm0.07$ & $92.58\pm0.06$ & $92.51\pm0.06$ & $92.43\pm0.07$ \\
%\hline
Reverse & $92.64\pm0.07$ & $92.60\pm0.06$ & $92.57\pm0.05$ & $92.49\pm0.06$ \\
\bottomrule
\end{tabular}
}
\label{table:4}
\end{center}
\end{minipage}

\end{table*}

\subsection{Discussion: the overwriting attack}
As a prospective application of capsulation attack in the field, it is worth discussing the timestamp issue.
To cope with the watermark overwriting attack, i.e., a malicious writes its identity information into a DNN and sells the model to confuse the ownership verification, it is necessary that the watermark be correlated with a timestamp.
Given the unforgeability of existing DNN watermarking schemes, such a correlation can be established by recording the ownership evidence (e.g., \texttt{Key} for ordinary backdoor-based watermarks and $\left\{\texttt{Key},\texttt{Merkle}(\mathbf{t}_{1},\mathbf{t}_{2},\cdots,\mathbf{t}_{N})\right\}$ for the reverse-backdoor scheme) into a distributed ledger.
This blockchain-based record adds a unique timestamp to the ownership evidence and hence the adversary conducting overwriting with a later timestamp can be accurately defeated.

However, under the ideal black-box setting, two malicious parties denoted as Alice and Bob can cooperate to succeed in copyright overwriting a third party, e.g., Carol's DNN, using the capsulation attack.
Concretely, Alice firstly announces its evidence pair $\left\{\texttt{Key}^{\text{A}},\texttt{Merkle}^{\text{A}}\right\}$ with underlying triggers $(\mathbf{t}^{\text{A}}_{1},\mathbf{t}^{\text{A}}_{2},\cdots,\mathbf{t}^{\text{A}}_{N})$ at time $t_{1}$ without engineering any model.
Then Alice and Bob steal Carol's product, whose watermark is correlated with time $t_{2}>t_{1}$.
Bob deploys the pirated DNN as its black-box service in a capsulation with
$$f(\mathbf{x})=\left\{\begin{aligned}
1,&\text{ }\exists n\in(1,2,\cdots,N),\mathbf{x}=\mathbf{t}^{\text{A}}_{n}\\
0,&\text{ otherwise}
\end{aligned} \right.,$$
and $l(\mathbf{x})$ returns the legal label consistent with $\texttt{Key}^{\text{A}}$ and $\texttt{Merkle}^{\text{A}}$.
Finally, if Carol declares its ownership of Bob's service then Alice can claim the ownership with an earlier timestamp and accuse Carol of breaching its intellectual property.
It is impossible to distinguish who is the authentic owner so the timestamp setting in the black-box setting is of no utility and the overwriting threat remains.
Using the reverse-backdoor scheme does not help in resolving this dilemma.
The previous discussion indicates that the following three requirements for DNN watermarking schemes cannot hold simultaneously 
(i) the black-box scenario
(ii) the availability of the watermarking scheme, and 
(iii) the unforgeable timestamp. 
Nonetheless, any two out of these three requirements can coexist in cases detailed in Table~\ref{table:5}. 
\begin{table}[!t]
\centering
\caption{The solutions to the different configurations of DNN watermark application requirements and the corresponding problems.  
(i) the black-box scenario
(ii) the availability of the watermarking scheme, and 
(iii) the unforgeable timestamp.}
\vspace {-2.5mm}
\begin{center}
\scalebox{0.85}{
\begin{tabular}{c|c|c|c|c}
\toprule
\textbf{(i)} & \textbf{(ii)} & \textbf{(iii)} & \textbf{Solution} & \textbf{Problem} \\
\hline
\checkmark & \checkmark & $\times$ & Current schemes & Overwriting attack \\
\hline
\checkmark & $\times$ & \checkmark & Secret algorithms & Easy to forge \\
\hline
$\times$ & \checkmark & \checkmark & White-box schemes & Limited scope \\
\bottomrule
\end{tabular}
}
\label{table:5}
\end{center}
\end{table}

\section{Conclusion and Future Work}
Due to the difference in purpose and threat model, backdoor attacks and backdoor-based DNN watermarking schemes have fundamental differences.
We claim that the ability to incorporate the owner's digital identity into the protected DNN makes backdoor-based DNN watermarks vulnerable to the capsulation attack. 
This idea has been empirically justified by our experiments.
To solve the threat of the capsulation attack while preserving the identity encoding's unforgeability, we propose a reverse backdoor-based DNN watermarking scheme and establish its advantage by analysis and experiments.

It is remarkable that the threats to copyright protection of DNN do not only involve the operations at the level of the neural network.
Flaws in assumptions of the scenario or the ownership verification protocol can easily make a secure watermarking scheme fragile and it is necessary to pay more attention to these aspects.

\vspace {5mm}
\bibliographystyle{IEEEbib}
%\centerline{\bf{References}}

\end{document}